\documentclass[twocolumn,floatfix,aps,prb,longbibliography]{revtex4-2}
\usepackage{graphicx}
\usepackage{dcolumn}
\usepackage{amsmath}
\usepackage{amssymb}
\usepackage{amsbsy}
\usepackage{units}

\begin{document}

\title{Externally controlled and switchable 2D electron gas at the Rashba interface 
 between ferroelectrics and heavy \texorpdfstring{$d$}{d} metals}

\author{T. Aull} \email{thorsten.aull@physik.uni-halle.de}
\affiliation{Institute of Physics, Martin Luther University Halle-Wittenberg,
             D-06099 Halle, Germany}

\author{I. V. Maznichenko}
\affiliation{Institute of Physics, Martin Luther University Halle-Wittenberg,
             D-06099 Halle, Germany}

\author{S. Ostanin}
\affiliation{Institute of Physics, Martin Luther University Halle-Wittenberg,
             D-06099 Halle, Germany}

\author{E. \c{S}a\c{s}{\i}o\u{g}lu}
\affiliation{Institute of Physics, Martin Luther University Halle-Wittenberg,
             D-06099 Halle, Germany}

\author{I. Mertig}
\affiliation{Institute of Physics, Martin Luther University Halle-Wittenberg,
             D-06099 Halle, Germany}

\begin{abstract}
 Strong spin-orbit coupling in noncentrosymmetric materials and interfaces 
 results in remarkable physical phenomena, such as nontrivial spin textures, which
 may exhibit Rashba, Dresselhaus, and other intricated configurations. This provides a
 promising basis for nonvolatile spintronic devices and further implications. 
 Here, we simulate from first principles a two-dimensional electron gas in 
 ultrathin platinum and palladium layers grown on ferroelectric PbTiO$_3$(001).
 The latter allows, in principle, to switch and control the spin-to-charge conversion 
 by the polarization reversal. We show how the band structure and its Rashba
 splitting differ in the Pt and Pd overlayers and how these electronic features change 
 with increasing the overlayer thickness and upon reversal of polarization.
 Besides, for both overlayers, we simulated their current-voltage ($I-V$) characteristics,
 the resistance of which upon the polarization reversal changes between 20~\% and several hundred percent.
 The reported findings can be used to model directly the Rashba-Edelstein effect.
\end{abstract}

\maketitle

\section{Introduction}
\label{sec:introduction}

 Reversal of polarization in ferroelectrics (FE) which is achieved by applying 
 an electric field allows to induce nonvolatile electrons or, alternatively, holes 
 into the material which overlays the FE substrates.~\cite{Lorenz-JPhysD-2016}
 So far, however, in the context of a two-dimensional electron gas (2DEG), research 
 focuses mostly on the TiO$_2$-terminated (001) surface of paraelectric SrTiO$_3$ (STO). 
 The 2DEG emerges there beneath epitaxially grown LaAlO$_3$ (LAO)~\cite{Ohtomo2004} or 
 similar polar perovskites~\cite{Li2015,our-AdvMater}.
 Presently, the nature of this 2DEG  and the role of polar overlayers are well understood
 (see Refs.~\onlinecite{Stemmer2014,MaznichenkoPSS2019} and references therein).
 LAO/STO(001), for instance, shows its 2DEG mobility which is one order of magnitude 
 higher than that of silicon based transistors. Moreover, the STO interface with some 
 defective oxides possesses 2D superconductivity~\cite{Reyren2007,hurand2015field,singh2018competition}
 and induced magnetism seen as anomalous Hall effect in magneto-transport 
 measurements.~\cite{Park2020NatComm}

 Alternatively, the 2DEG emerges in STO(001) by the deposition of ultrathin film of Al
 or some other reactive metals.~\cite{santander2011two,brehin2020switchable,vicente2021metal}
 The metal overlayer partially pulls out oxygen 
 from STO and transforms into binary oxide. Meanwhile, the upper 1-nm-thick layers of STO 
 become oxygen-deficient and marginally polar. In this scenario, the broken inversion symmetry
 at the 2DEG interface lifts spectral degeneracy that is
 known as the effect of Rashba spin-orbit coupling (SOC).~\cite{BychkovRashba}
 For instance, the Al/STO(001) shows at 2~K the electric-field-induced switchable polarization
 of $\sim$4$\mu$C cm$^{-2}$, while the 2DEG exhibits a sizable Rashba SOC~\cite{CavigliaPRL2010}
 with relatively high conversion efficiency.~\cite{vaz2019mapping}

 A spin-orbitronic concept of the FE-controlled spin–to.charge conversion has been 
 suggested recently by P. No{\"e}l \textit{et al.}~\cite{Noel2020} who demonstrate this 
 phenomenon experimentally using NiFe(20~nm)/Al(0.9~nm)/STO.~\cite{Noel2020}
 
The authors show how the nonmagnetic Rashba system efficiently generates 
spin current from charge current due to the Rashba-Edelstein effect~\cite{Edelstein1990}.
The breaking of inversion symmetry at the interface results in the out-of-plane electric field. 
Then, the in-plane charge current via interfacial 2DEG produces a transverse spin density, 
which can diffuse as a spin current into the adjacent material~\cite{Kondou_NatPhys2016}. 
Conversely, the injection of a spin current into the Rashba state will produce the charge 
current due to the inverse Edelstein effect~\cite{Sanchez_NatCommun2013}.
Most importantly, the FE polarization reversal changes the sign of the local
electric field that reverses the chirality of the spin textures in the both Rashba-split
Fermi contours (see Fig.~1 of Ref.~\onlinecite{Noel2020}). On the other hand,
the charge current sign, generated through the inverse Edelstein effect,
should depend on the FE polarization state. This mechanism offers the basis for
the bipolar memory device and other logic devices.~\cite{ManipatruniNature2019}

 This work was motivated by the interconverting charge and spin currents through 
 the direct and inverse Edelstein and spin Hall 
 effects.~\cite{lesne2016highly,vaz2018oxide,vaz2019mapping,Noel2020,trier2020electric,vaz2020determining,dang2020ultrafast,johansson2021spin}
 When the robust ferroelectric substrate, such as PbTiO$_3$(001) (PTO), is used 
 instead of the STO 2DEG, 
 this makes possible to extend the ordinary electric-field dependence of Rashba 
 SOC~\cite{CavigliaPRL2010,MirhosseiniPRB2010} to its switching option since 
 the polarization reversal may accumulate/deplete 2DEG electrons. 
 Highly reactive metal is not needed now for the overlayer material. Here, 
 we simulated from first principles the 2DEG in the few-monolayer-thick 
 nonmagnetic metals ($Me$): Pd and Pt. 
 In this scenario, the formation of 2DEG moves from the two upper layers 
 of STO into the metal overlayer. Therefore, the picture of Rashba SOC changes.
 In the O deficient ABO$_3$ perovskite layer, the Fermi level crosses the bottom 
 B-cation conduction band, which is splitted by SOC into two parabolas around 
 the $\Gamma$ point of the Brillouin zone.
 In the case of $Me$/PTO ($Me$ $=$ Pd, Pt), its 2DEG is formed by multiple $Me$
 $d$-branches that is typical for the heavy d-metal and broken inversion symmetry.

It should be noted that the interfaces between perovskites and metals provoke the formation
of oxygen vacancies, oxide layers and the intermixing of cations across the
interface. The one-unit-cell steps can be also seen there.~\cite{Meyerheim2011}
It is established now that perovskites grow in complete unit cells while their
(001) surfaces are usually TiO$_2$-terminated. To obtain the PbO-terminated PTO,
a special procedure is required.~\cite{guo2021interface}
The interfacial defects in multiferroic tunnel junctions reduce the functional
insulating thickness of the barriers~\cite{quindeau2015origin} whereas the functionality 
and weakly degrading properties of electrodes are not characterized. 
The issue of PTO(001) passivated by Pb was discussed in Ref.~\onlinecite{kurasawa2005surface}.
Using the angle-resolved x-ray photoemission spectroscopy, the authors found that
metallic Pb diffuses into the Pt layer during the Pt deposition on PTO(001) thin films.
As the result, a defective layer at the Pt/PTO(001) interface produces the observed Fermi
 energy pinning. Obviously, the Pt overlayers diluted by Pb retain many properties 
 of ideal material, including the screening of the FE dipole. Since Pb is a heavy chemical element,
 the Pb substitute should not affect the Rashba SOC significantly. We relaxed 
 the chemically perfect and TiO$_2$-terminated interfaces of metal/PTO only.
 In principle, to obtain the spin-to-charge conversion the nonmagnetic-metal/PTO system 
 needs to be covered by an extra layer of NiFe. Here, we do not model 
 the effect of proximity to the ferromagnetic layers.
 
Focusing on the Rashba SOC in 2DEG of dually polar Pd/PTO and Pt/PTO, we calculated and 
 compared their band structures. Besides, the distinguished Rashba splittings 
 are evaluated. Finally, electroresistance and its dependence on the polarization reversal
 as well as the $Me$ overlayer thickness are presented that was not performed so far.
 We anticipate that the relative
 resistance, calculated as {$[R$(P$\uparrow$) - $R$(P$\downarrow$)$] / R$(P$\downarrow)$},
 changes, depending on the number of metallic overlayers, between 20~\% and more
 than 400~\%. Thus, the findings reported  may attract extensive attention.

 \begin{figure*}
        \includegraphics[width=0.88\textwidth]{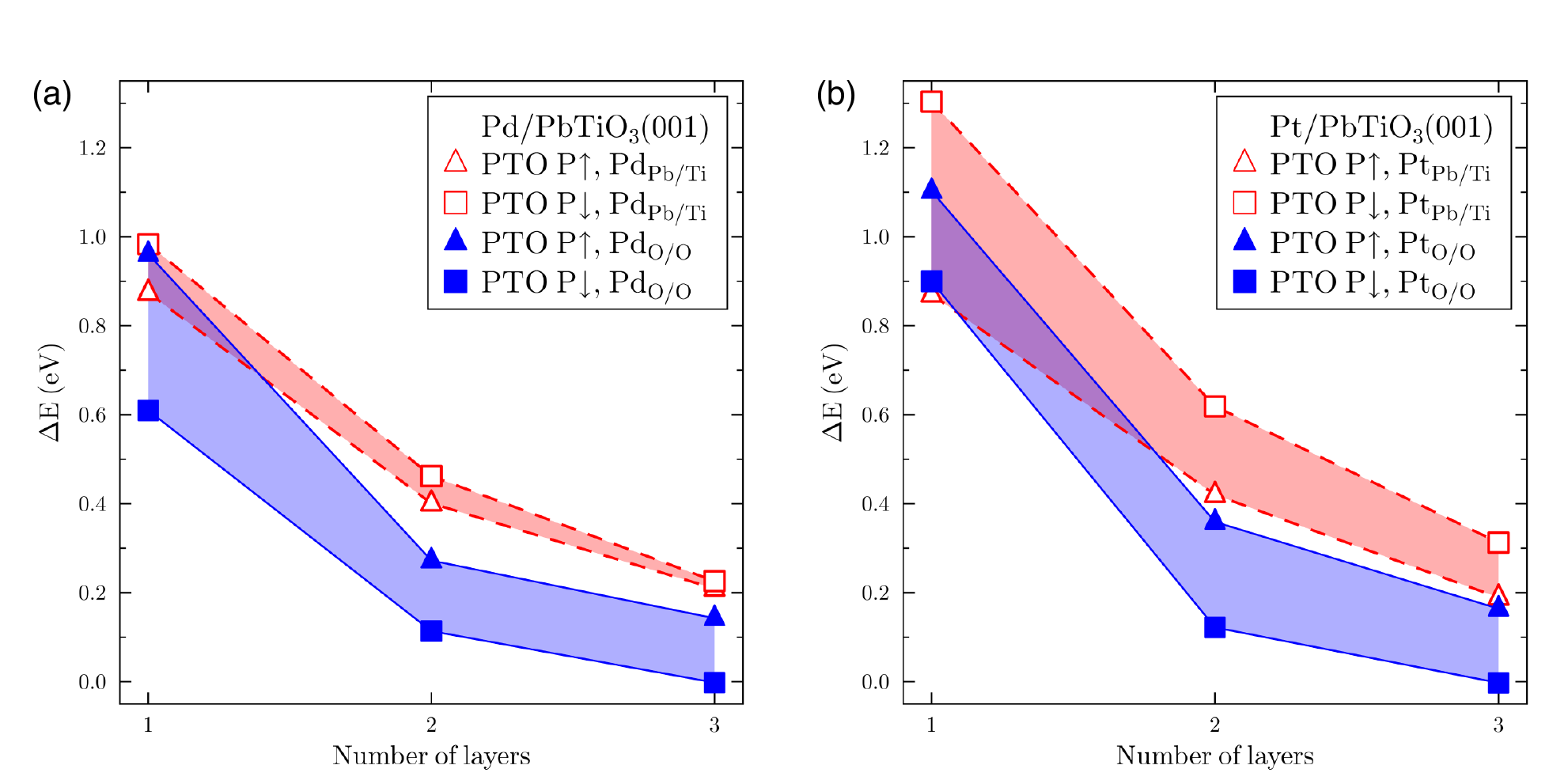}
        \caption{The relative energy difference (in eV per one $Me$) of (a) $L$ $\cdot$
         (Pd$_2$)/PTO~({\bf P})  and (b) $L$ $\cdot$ (Pt$_2$)/PTO~({\bf P})
 relaxed for each {\bf P}, each $L$ $=$ 1, 2, 3 and two interfacial configurations
 on top of oxygens (O/O) and on top of cations (Ti/Pb). For each {\bf P} and
 $Me$-configuration, the energetics is shown as a function of $L$ relatively chosen zero
 which corresponds to the lowest-energy configurations (O/O) for $L$ $=$ 3.}
        \label{fig1}
\end{figure*}

\section{Details of calculations}
\label{sec:details}

In all {\it ab initio}  calculations performed here, the TiO$_2$-terminated PTO(001) 
was modeled using a reliable set-up, within the slab geometry with a vacuum
layer.~\cite{fechner2008effect}
Over the past decade, the interfaces between the  TiO$_2$-terminated perovskites
and metals were studied in the context of multiferroic tunnel junctions,
in which the FE barrier is placed between two ferromagnetic electrodes 
that enables four distinct resistance states.~\cite{Garcia-Science2010,Pantel2012}
That is why the ferromagnetic 3$d$ metals, mainly, and some other ferromagnetic compounds 
adjusted as electrodes were simulated.~\cite{duan2006predicted,Meyerheim2011,velev2009magnetic,quindeau2015origin,fechner2008magnetic,fechner2010ab,borek2012first,borisov2014magnetoelectric,borisov2015spin,liu2016enhanced,borisov2017multiferroic,li2017polarization,umeno2009ab} 
Curiously, despite relatively strong magneto-electric coupling detected for all these 
FE/ferromagnetic interfaces, the robustly switched magnetic order of electrodes due to 
the barrier polarization reversal was not evidenced until now. 
 
Regarding the nonmagnetic electrode materials, the 4$d$ and 5$d$ f.c.c. metals represent 
the most suitable option. In particular, the lattice mismatch between PTO and Pd (Pt) is 
less than 0.7~\% (1.5~\%). The Pt/BaTiO$_3$ interface is well studied experimentally and 
theoretically from first principles.~\cite{velev2007effect,meyerheim2013tuning,yang2020role}
As for Pd/PTO and Pt/PTO, so far only the latter interface was simulated 
from first principles.~\cite{al2010density} However, the structure  
changes and, especially, the changes in the 2DEG, occurring upon the PTO polarization 
reversal at the presence of Rashba SOC, need a more detailed consideration.

To treat the interplay between electronic and structural properties of 
ferroelectric PTO(001) and ultrathin Pt(Pd) overlayers, we used
the two density functional theory (DFT) packages: 
the Vienna \textit{Ab initio} Simulation Package (\textsc{VASP})~\cite{VASP} 
and \textsc{QuantumATK} package~\cite{QuantumATK}. 
The idea of using these different codes for the same material is not to evaluate 
their accuracy, although we keep in mind this double-check. The \textsc{VASP} has the plane-wave 
basis set, which provides more accurate energetics and structural optimization, while
\textsc{QuantumATK} allows to simulate the semi-infinite supercell. It is important to demonstrate 
that the set up of finite and relatively thin supercell can mimic well the ferroelectric 
PTO(001). Here, we performed the test calculation using this semi-infinite option of 
\textsc{QuantumATK}. The second reason of using \textsc{QuantumATK} is its transport implementations
which allow to simulate the electroresistance in $Me$/PTO.

 The geometric relaxations were obtained by \textsc{VASP}. 
The thickness of metallic overlayers varies between one and three monolayers (ML)
with two Pt(Pd) per
ML. All these metallic atoms as well as atoms of 
the 2-u.c.-thick PTO near the interface were allowed to relax. Beneath that, to mimic
semi-infinite FE, we kept the 3-u.c.-thick PTO fixed, optimized already 
for each of its two directions of polarization.     
The electron-ion interactions within \textsc{VASP} calculations were described by the 
projector-augmented wave pseudopotentials and the electronic wave functions were
represented by plane waves with an energy cutoff of 450~eV.
 Although we focus here on the Perdew-Burke-Ernzerhof (PBE) generalized-gradient
 approximation (GGA)~\cite{Perdew1996} to the exchange-correlation potential, 
 the local density approximation (LDA) within DFT, was used as well.
 The P$\downarrow$ and P$\uparrow$ configurations of $Me$/PTO were relaxed separately 
 for each overlayer material and for each thickness $L$ $=$ 1,2,3.
 Ionic relaxation in the Me overlayers and layers of PTO beneath the interface
 was performed using the conjugate-gradient algorithm until the Hellmann-Feynman forces 
 became less than $1 \times 10^{-2}$~eV/{\AA}. The use of the $6 \times 6 \times 2$ 
 {\bf k}-mesh yielded the reliable atomic positions.
 The density of states (DOS) was obtained then using the tetrahedron method on
 the $\Gamma$-centered and compacted {\bf k}-mesh with no smearing for the electronic
 occupations.
 A cross-check of the interface structure and Rashba SOC on the electronic states 
 was done using the \textsc{QuantumATK} package. Thus, various computed quantities 
 were carefully compared among the two DFT codes to obtain consistent results while 
 their reliability was achieved by numerous convergence tests.
 
 To receive the layer-resolution of the DOS and band structure, we used the \textsc{QuantumATK} package which uses linear combinations of atomic orbitals (LCAO) as basis set.
 For all of these calculations we used the slab configuration optimized with \textsc{VASP} and applied a $15 \times 15 \times 3$ $\Gamma$-centered \textbf{k}-point mesh and a density-mesh cutoff energy of 280~Ry and a broadening of 25~meV together with fully-relativistic norm-conserving PseudoDojo pseudopotentials.~\cite{QuantumATKPseudoDojo}
Within this setup we calculated the fat bands and projected the DOS on each layer.

\begin{figure}[!b]
		\includegraphics[width=0.49\textwidth]{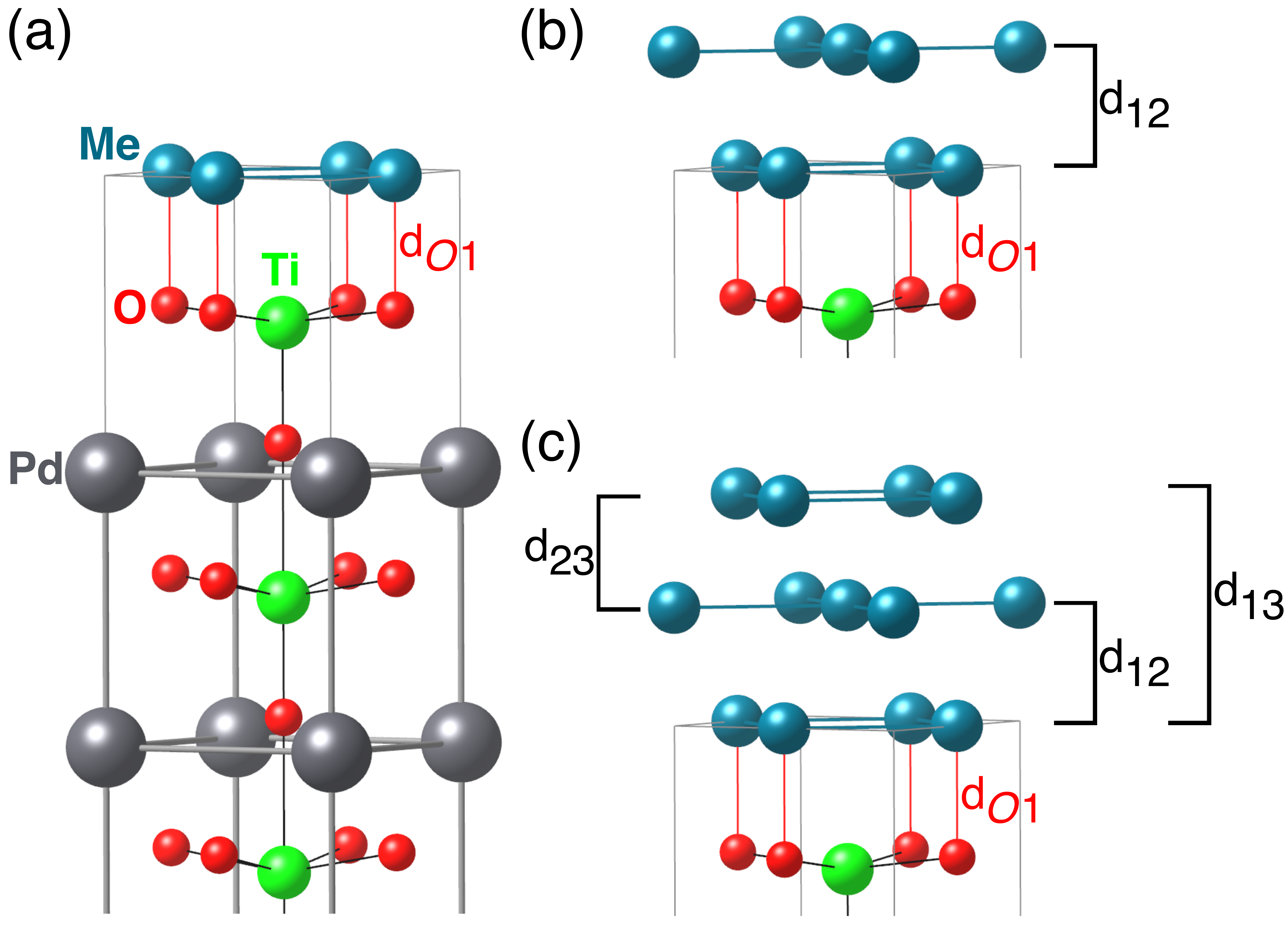}
		\caption{Relaxed structures of $L$ $\cdot$ $Me$/PbTiO$_3$ ($Me$ $=$ Pd, Pt) are shown 
 for $L$ $=$ 1 (a), $L$ $=$ 2 (b) and $L$ $=$ 3 (c). This corresponds to the PTO polarization P$\downarrow$ while
 all interlayer distances (d$_{ij}$) are given in Tab.~\ref{tab1} for P$\downarrow$ and P$\uparrow$. 
 $Me$ atoms are represented in blue, Pb in grey, Ti in green, and O in red. 
 For $L$ $=$ 1, beneath the interfacial TiO$_2$ layer of PTO we display also its two unit cells after
 relaxation.}
	\label{fig2}
\end{figure}

\section{Results and discussion}

\begin{figure*}
\centering
    \includegraphics[width=0.85\textwidth]{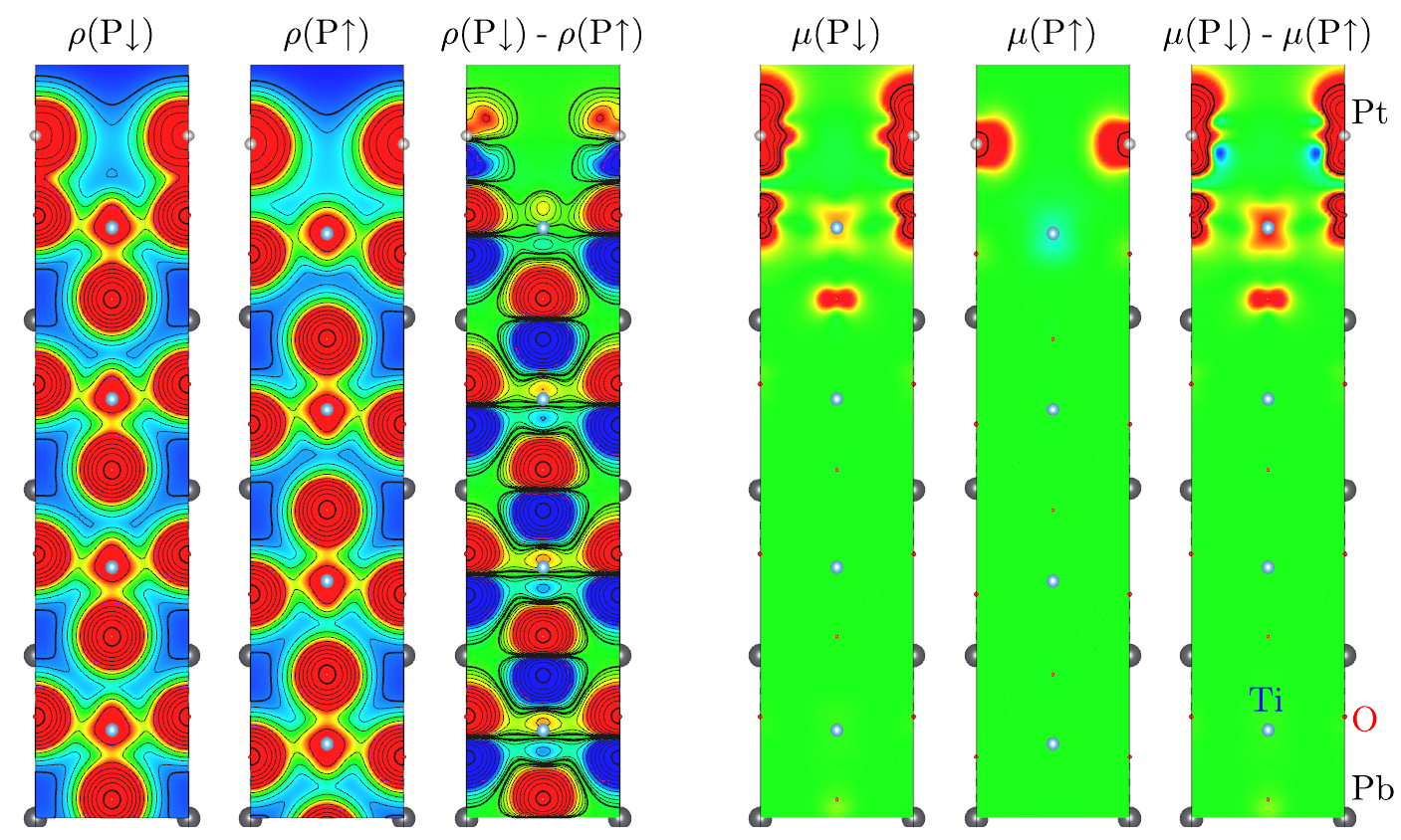}
    \caption{The charge densities $\rho$ (left) and magnetization densities $\mu$ (right) of 1ML Pt on the P$\downarrow$/P$\uparrow$-poled  PTO(001) as well as their differences  $\rho$(P$\downarrow$)--$\rho$(P$\uparrow$) and $\mu$(P$\downarrow$)--$\mu$(P$\uparrow$) (right) calculated upon reversal of $\bf P$. 
    The density cut [0yz] through the Pt atoms is shown, while color scales in 
    arbitrary units.} 
    \label{charge}
\end{figure*}
 
\subsection{Structural properties and induced extra carriers}

Here we focus on relaxed structure of the metal ($Me$ $=$ Pt, Pd) MLs.
The outermost and key TiO$_2$ layer of PTO(001) is denoted S, while 
the intralayer $z$-displacements in each PTO layer, $\delta = z(cation) - z(O)$, 
are positive (negative) for P$\uparrow$ (P$\downarrow$). 
The $Me$ layers, which form tetragonally distorted f.c.c lattice, 
are indicated as S+1, S+2  and S+3. In the first $Me$ layer, S+1,
its two sites were relaxed either on top of the O sites (O/O) or
on top of the PTO cation sites (Ti/Pb).
 
In Figs.~\hyperref[fig1]{1\,(a)} and \hyperref[fig1]{1\,(b)}, the energetics
of relaxed $L$ $\cdot$ (Pd$_2$)/PTO~({\bf P}) and $L$ $\cdot$ (Pt$_2$)/PTO~({\bf P}) 
is shown, respectively, as a function of the $Me$ thickness $L$ $=$ 1, 2, 3. 
For each case of the FE polarization {\bf P}
we subtracted the energy of relaxed substrate from the total energy of the system
and then normalized the results per one $Me$ atom.
The lowest normalized energy, which corresponds to the (O/O) configuration of
$L$ $=$ 3 and {\bf P} $=$ P$\downarrow$, was selected as the zero energy.
Thus, Fig.~\ref{fig1} represents along its $y$-axis the relative 
differences in energy, as compared to that of the lowest normalized difference.
We found that the interfacial (O/O) configuration, with both $Me$ atoms on top of 
oxygens of the  P$\downarrow$-poled PTO(001), is strongly energetically preferable, except 
one competitive case of 1-ML-thick Pt on PTO(P$\uparrow$), where both Pt can sit above 
cations in configuration (Ti/Pb). The latter case is clearly seen in 
Fig.~\hyperref[fig1]{1\,(b)}. However, by placing the S+2 and S+3 layers, the
(O/O) configuration  becomes again notably preferable. Surprisingly, for the 
(Ti/Pb) interfacial configurations of the two $Me$ overlayers whatever their thickness is, 
the polarization state P$\uparrow$ is favourable that contradicts to the configuration (O/O).  
For the latter, the P$\downarrow$ state of PTO is always favourable. Therefore, in the following 
we show and discuss the results obtained for the dually poled (O/O) configurations only.
In general, by analyzing the results shown in Fig.~\ref{fig1}, we 
find that the both systems Pd/PTO and Pt/PTO become energetically more stable for the thicker 
metallic overlayer, which obeys gradually the f.c.c. lattice starting from $L$ $=$ 3.

The $Me$ structures after relaxation are given in Fig.~\ref{fig2}.

We introduced there the interlayer structure parameters $d_{ij}$ and collected all computed 
data in Tab.~\ref{tab1} for the two states of poled PTO and each $L$.
For $L$ $=$ 1, the interfacial Pd--O (Pt--O) distance, d$_{O1}$ of 2.01~{\AA} (2.00~{\AA}),
which is obtained for P$\downarrow$, increases significantly upon polarization reversal 
to the value of 2.32~{\AA} (2.61~{\AA}). With increasing $L$ $>$ 1, the  P$\downarrow$-poled  
d$_{O1}$ increases slightly by 0.03--0.05~{\AA} for both metals, whereas the ${\bf P}$
reversal to P$\uparrow$ diminishes d$_{O1}$ by about 0.1~{\AA} and 0.3~{\AA} for
Pd and Pt, respectively.
     
For $L$ $=$ 2 and $Me$ $=$ Pd (Pt), the interlayer distance between S+1 and S+2, d$_{12}$, 
increases from 1.92~{\AA} (1.97~{\AA}) by 0.02~{\AA} (0.04~{\AA}) when  ${\bf P}$ changes from
P$\downarrow$ to P$\uparrow$. Growing the third $Me$ layer results in the 0.04~{\AA}
increase of d$_{12}$ that is seen for both metals and each ${\bf P}$. Meanwhile,
the separation S+2 and S+3, d$_{23}$, shows the systematic and notable decrease 
of 0.07~{\AA}, as compared to d$_{12}$. 
For $L$ $=$ 3, the interlayer separation d$_{13}$ between S+1 and S+3 gives
an idea of the $c$ lattice parameter. Since the in-plane lattice parameter of 3.88~{\AA} 
is fixed by the PTO substrate, we find that the first Pd  unit cell is cubic while
Pt is tetragonally elongated.

\begin{table}[b]
    \caption{Selected interlayer distances (in angstroms) of Pd/PTO and Pt/PTO after relaxation.
  The distance notation corresponds to Fig.~\ref{fig2}.}
    \label{tab1}
    \begin{ruledtabular}
    \begin{tabular}{@{}l*{8}{c}@{}}
        $Me$ & $\bm P$ & d$_{O1}$ & d$_{O1}$ & d$_{O1}$ & d$_{12}$ & d$_{12}$ & d$_{23}$ & d$_{13}$ \\
         & & (1ML) & (2ML) & (3ML) & (2ML) & (3ML) & (3ML) & (3ML)\\
    \hline
        Pd & P$\downarrow$ & 2.01 & 2.04 & 2.03 & 1.92 & 1.96 & 1.89 & 3.86\\
        Pd & P$\uparrow$ & 2.32 & 2.23 & 2.20 & 1.94 & 1.98 & 1.91 & 3.88\\
        Pt & P$\downarrow$ & 2.00 & 2.04 & 2.05 & 1.97 & 2.03 & 1.96 & 3.98\\
        Pt & P$\uparrow$ & 2.61 & 2.32 & 2.33 & 2.01 & 2.02 & 1.95 & 3.97\\
    \end{tabular}
    \end{ruledtabular}
\end{table}

\begin{figure}[!t]
	\includegraphics[width=0.465\textwidth]{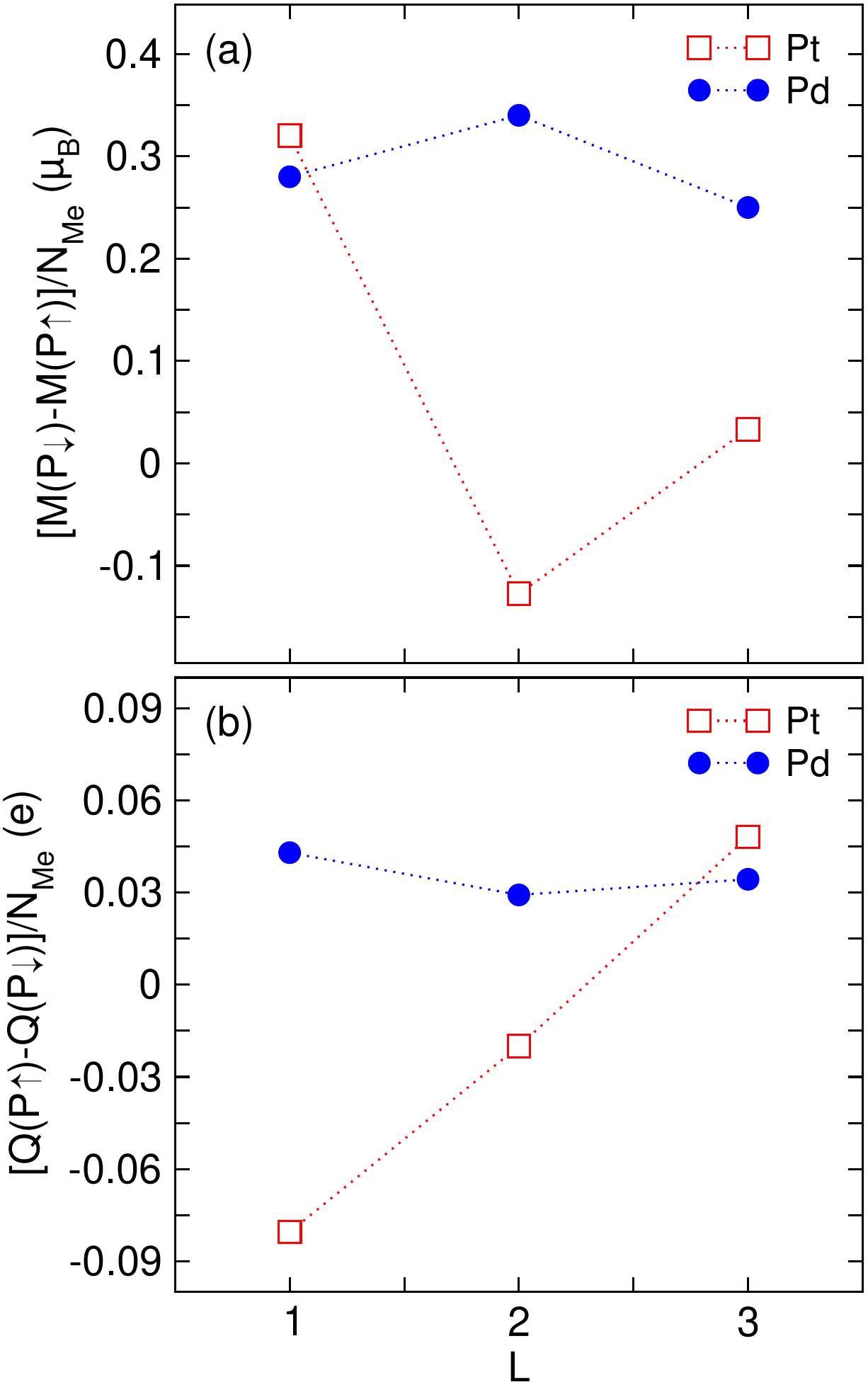}
	\caption{The difference in the induced magnetic moments (a) and orbital charges (b) 
 of $Me$ $=$ Pt, Pd between the PTO polarizations P$\uparrow$ and P$\downarrow$, which are 
 plotted as a function of the $Me$ thickness}
	\label{fig4}
\end{figure}

The changes in structure of $Me$/PTO, which occur due to the  ${\bf P}$-reversal,  
redistribute charges in the system. This is illustrated for the case $L$ $=$ 1 and $Me$ $=$ Pt 
in Fig.~\ref{charge} 
where the charge density cut, $\rho$, is plotted for each direction of ${\bf P}$ 
as well as the charge differences $\rho$(P$\downarrow$)--$\rho$(P$\uparrow$).
 To compare the two ${\bf P}$ densities and to plot the charge (magnetization) density 
 difference in the third panel of Fig.~\ref{charge}, the positions of Pb in the 
 layer (S-1) were coincided for P$\downarrow$ and P$\uparrow$.
The charges change dramatically in the PTO layers that is not surprising.
However, the interface (S) and Pt layer (S+1) display also the charge redistribution.   
To evaluate this effect, we show in the lower panel of Fig.~\ref{fig4}
the $Me$ charge difference, calculated between the PTO polarizations P$\uparrow$ 
and P$\downarrow$ and plotted  as a function of L. For $Me$ $=$ Pd and for all L, the state
P$\uparrow$ induces $\sim$0.04 electron per each Pd, as compared to P$\downarrow$.
The case of Pt is much more specific. For 1ML Pt, the P$\downarrow$ state of PTO creates
$\sim$0.08 electron more than P$\uparrow$. Then, with increasing L, the disbalance
between P$\uparrow$  and P$\downarrow$ becomes marginal for $L$ $=$ 2 and, finally, for $L$ $=$ 3
P$\uparrow$ overcomes with 0.05 electron, i.e. the value similar to that of 3ML Pd.

\begin{figure}[!t]
        \includegraphics[width=0.45\textwidth]{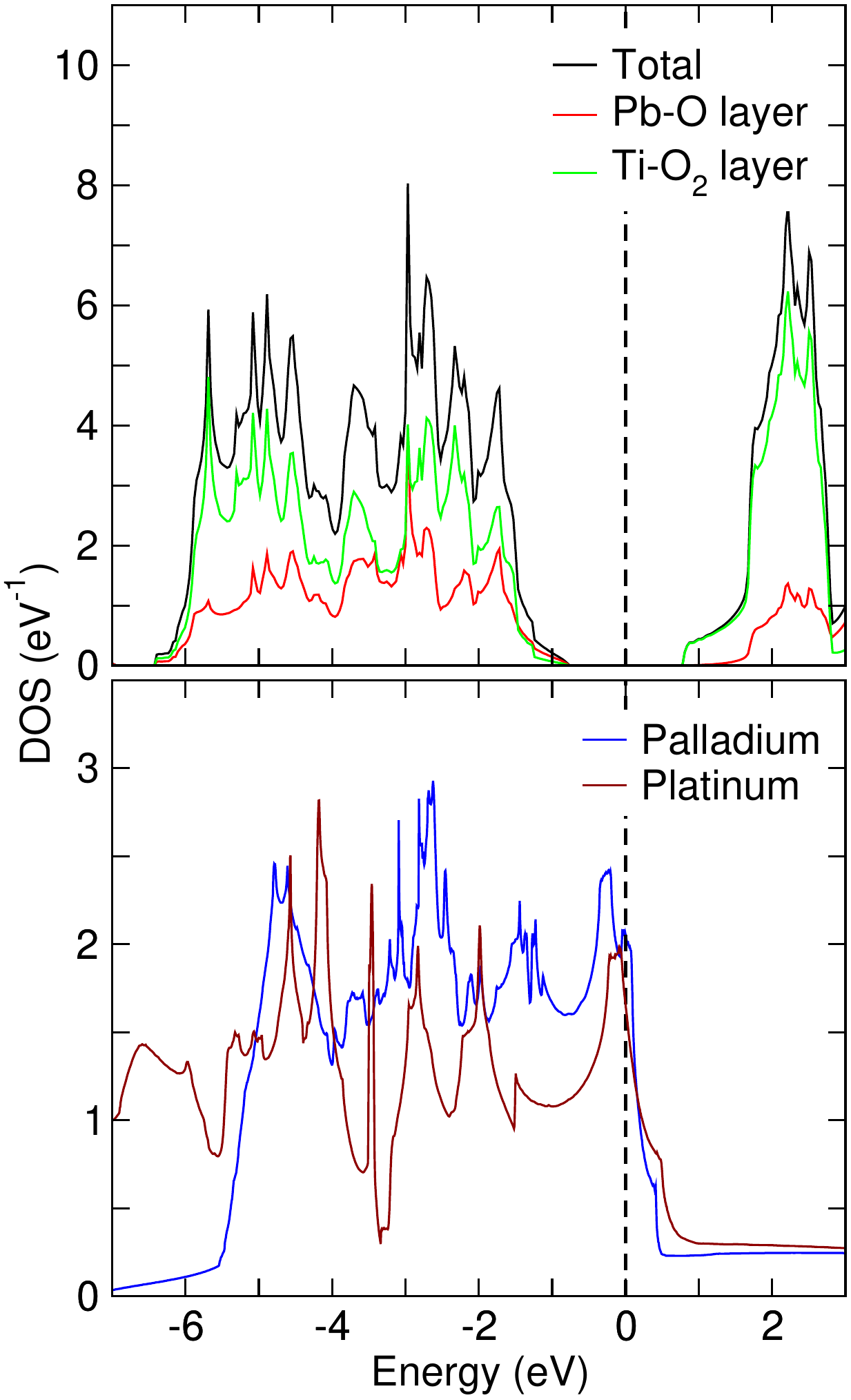}
        \caption{The total DOS of PbTiO$_3$ and its layer-resolved Pb-O (red) and Ti-O$_2$ (green) contributions are shown in the upper panel. 
              The lower panel shows the DOS of bulk Pd and Pt with respect to $E_F = 0$.}
        \label{fig5}
\end{figure}

The thin film geometry of $Me$ overlayer and extra electrons induced by PTO result 
in the magnetization densities, which are seen in  Fig.~\ref{charge} for dually
polar 1ML Pt together with their difference cut $\mu$(P$\downarrow$)--$\mu$(P$\uparrow$).
The $Me$ magnetic moments $m$ were calculated using \textsc{VASP} for each $L$. 
For $Me$ $=$ Pd, the P$\uparrow$ polarization  results in relatively small $m$,
which vary from 0.03 $\mu_B$ in the topmost layer of each $L$ to 0.18 $\mu_B$.
The $\bf P$-reversal induces the larger Pd moments which exceed 0.4 $\mu_B$ when $L$ $>$ 1.     
Since extra electrons fill in the minority spin band of the $Me$ d-states, the 
corresponding charge increase reduces the value of induced magnetic moment. 
The case of $Me$ $=$ Pt shows rather similar trend of induced magnetism, except 
Pt bilayer at P$\downarrow$ when all $m$ become marginal. This is because of 
relatively short interlayer separation that was discussed for iron bilayer
on PTO.~\cite{fechner2010ab}
   
It should be noted that the targeting spin-to-charge conversion needs the next
and robust ferromagnetic layer which covers nonmagnetic Me. That coverage may
seriously reduce or even suppress the $Me$ moments. Nevertheless, the scenario of
switchable induced magnetization can be simulated here, in absence of
ferromagnetic overlayer.   
In the top panel of Fig.~\ref{fig4}, we plot the $Me$ moment difference between 
P$\uparrow$ and P$\downarrow$ as a function of $L$. As one can see there,
Pd represents a reliable switch, indeed, whereas Pt seems less attractive 
because of the case of weakly magnetic $L$ $=$ 2. 

\begin{figure}[!t]
        \includegraphics[width=0.45\textwidth]{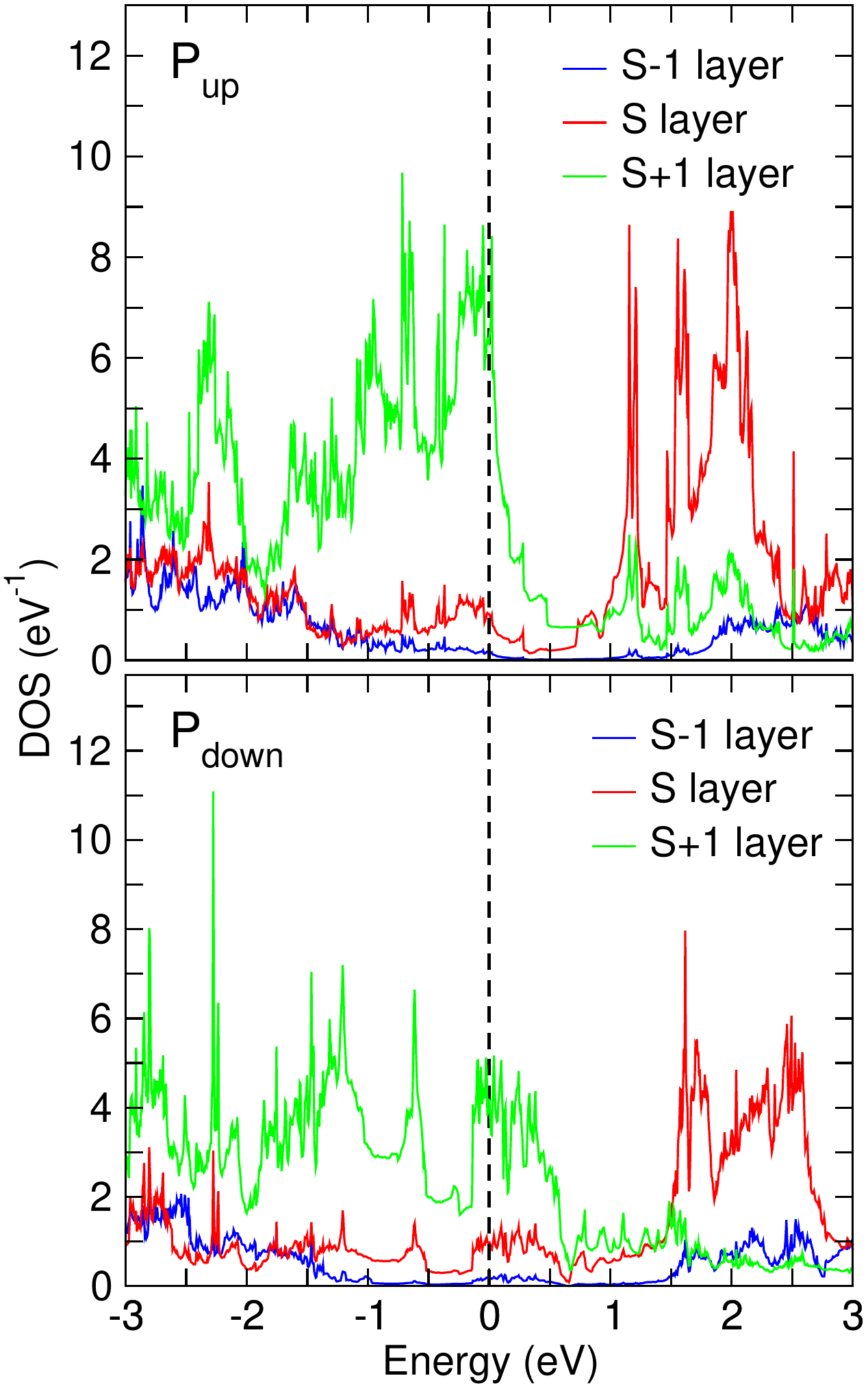}
        \caption{The layer resolved DOS of 1ML Pt/PTO(001) calculated for the two PTO polarizations P$\uparrow$
        (top) and P$\downarrow$ (bottom). The DOS contributions from the Pt layer S+1 are shown by the green line,
         while blue and red lines show the DOS of the PTO layers S$-$1 and S, respectively.}
        \label{fig6}
\end{figure}

\begin{figure}[!t]
    \center
	\includegraphics[width=0.49\textwidth]{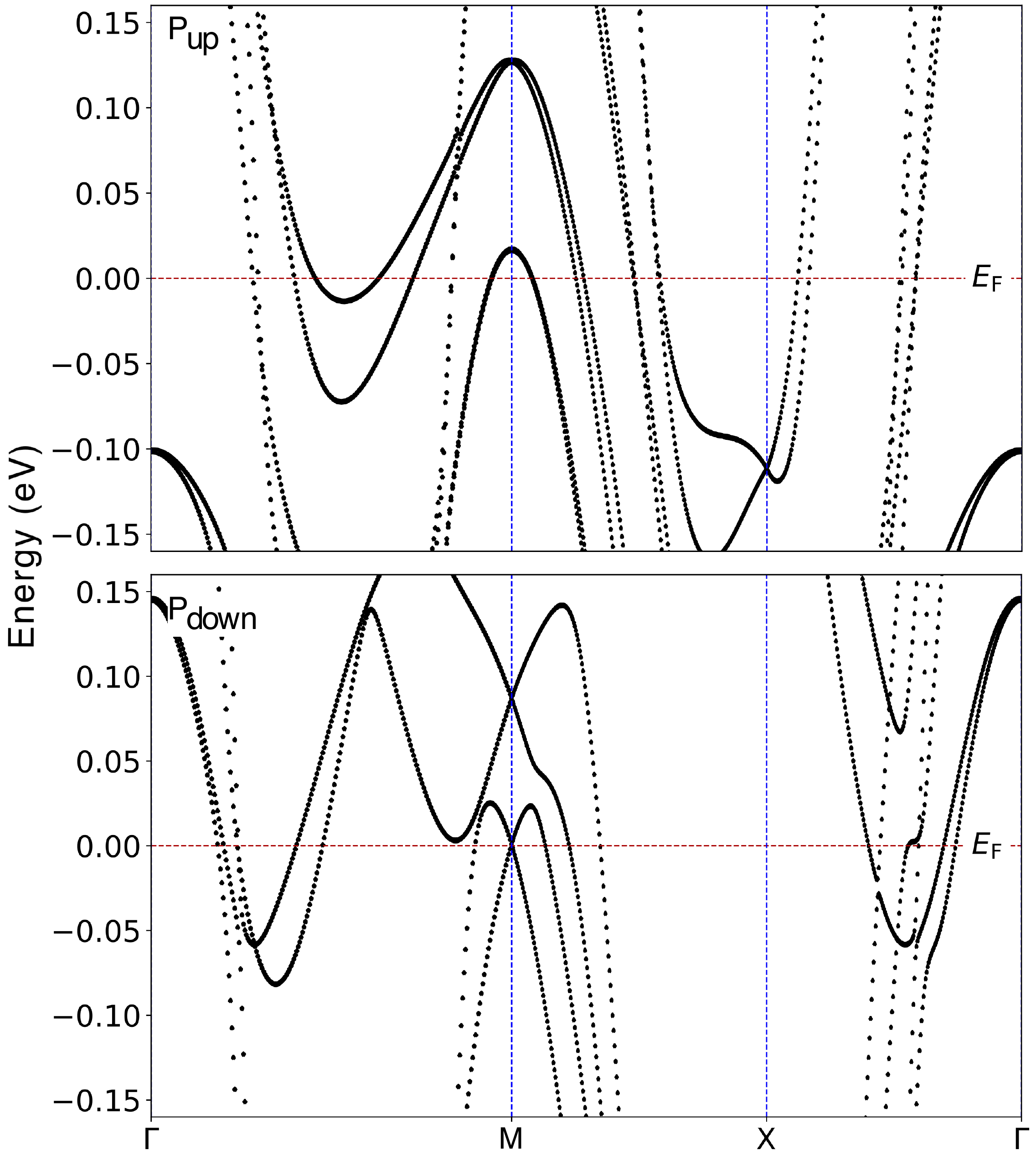}
	\caption{Surface band structure of 1ML Pt/PTO calculated for polarization 
     P$\uparrow$ (P$\downarrow$) shown in the upper (lower) panel.}
	\label{fig7}
\end{figure}

\begin{figure*}
    \center
	\includegraphics[width=0.95\textwidth]{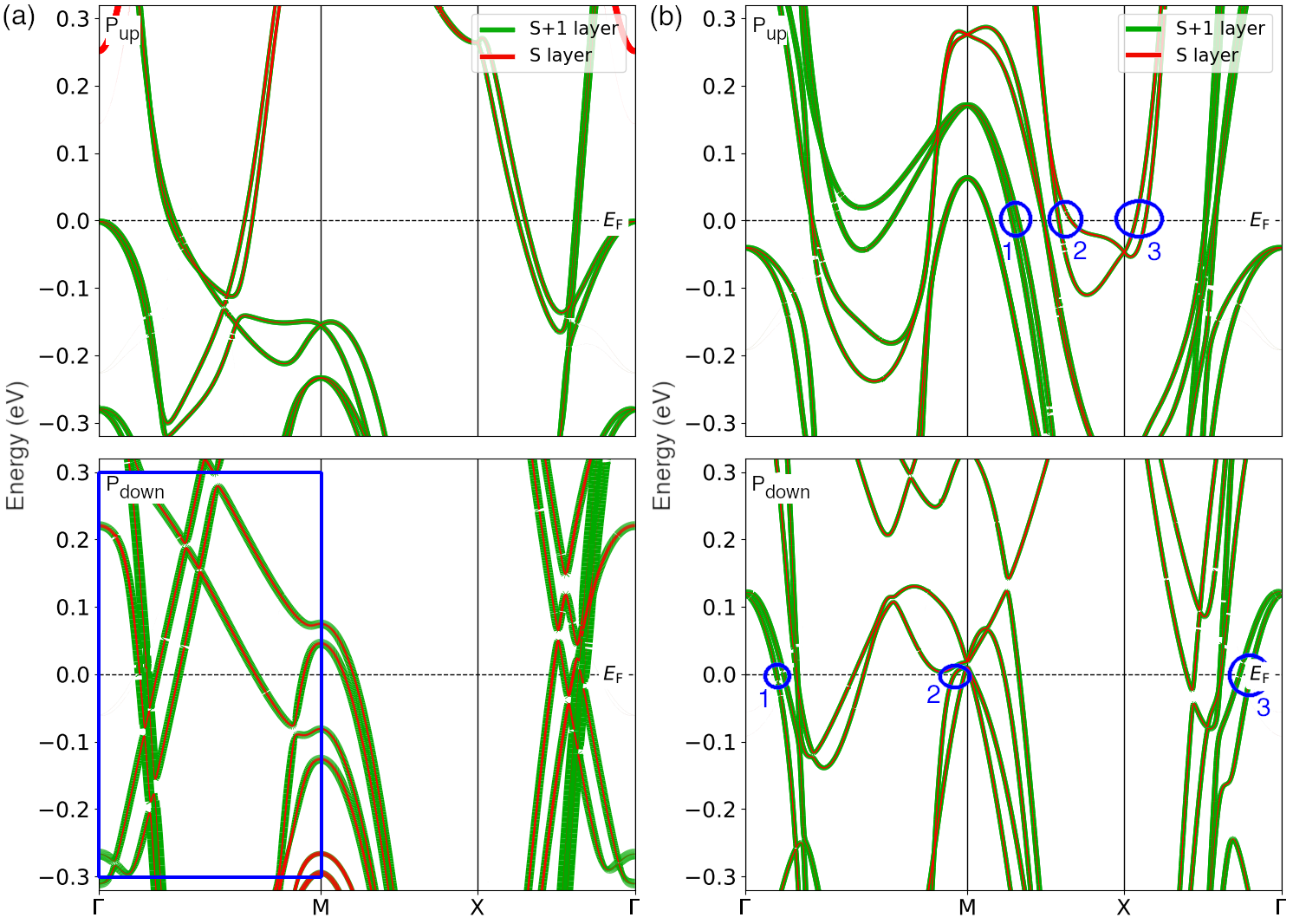}
	\caption{Band structure of (a) 1ML Pd/PTO and (b) 1ML Pt/PTO calculated for polarization P$\uparrow$ (P$\downarrow$) shown in the upper (lower) panel. 
	Red and green lines denote the bands emerging from the layers S and S+1, respectively. The linewidth is proportional to the state density $n_i (E)$ of each branch at given energy.
	The band structure within the blue rectangle is displayed spin-resolved in Fig.~\hyperref[fig10]{10\,(b)} while the enumerated blue areas mark some selected Rashba splittings near $E_F$, the details of which are collected in Table~\ref{tab2}.}
	\label{fig8}
\end{figure*}

\subsection{2DEG and its band structure}

\begin{table}[!b]
\caption{The Rashba SOC $k$-splittings ($\Delta k$) calculated at $E_F$ for 1ML Pt/PTO.
 The selected $\Delta k$ for each case of polarization are shown and
 enumerated in Fig.~\hyperref[fig8]{8\,(b)}.}
\label{tab2}
    \begin{ruledtabular}
    \begin{tabular}{cc}
        Position & $\Delta k$  \\
        & (1/{\AA})  \\
        \hline
        P${\uparrow,1}$ & 0.031  \\
        P${\uparrow,2}$ & 0.070  \\
        P${\uparrow,3}$ & 0.052  \\
        \hline
        P${\downarrow,1}$ & 0.053 \\
        P${\downarrow,2}$ & 0.079 \\
        P${\downarrow,3}$ & 0.057  
    \end{tabular}
    \end{ruledtabular}
\end{table}

\begin{figure*}
	\includegraphics[width=0.95\textwidth]{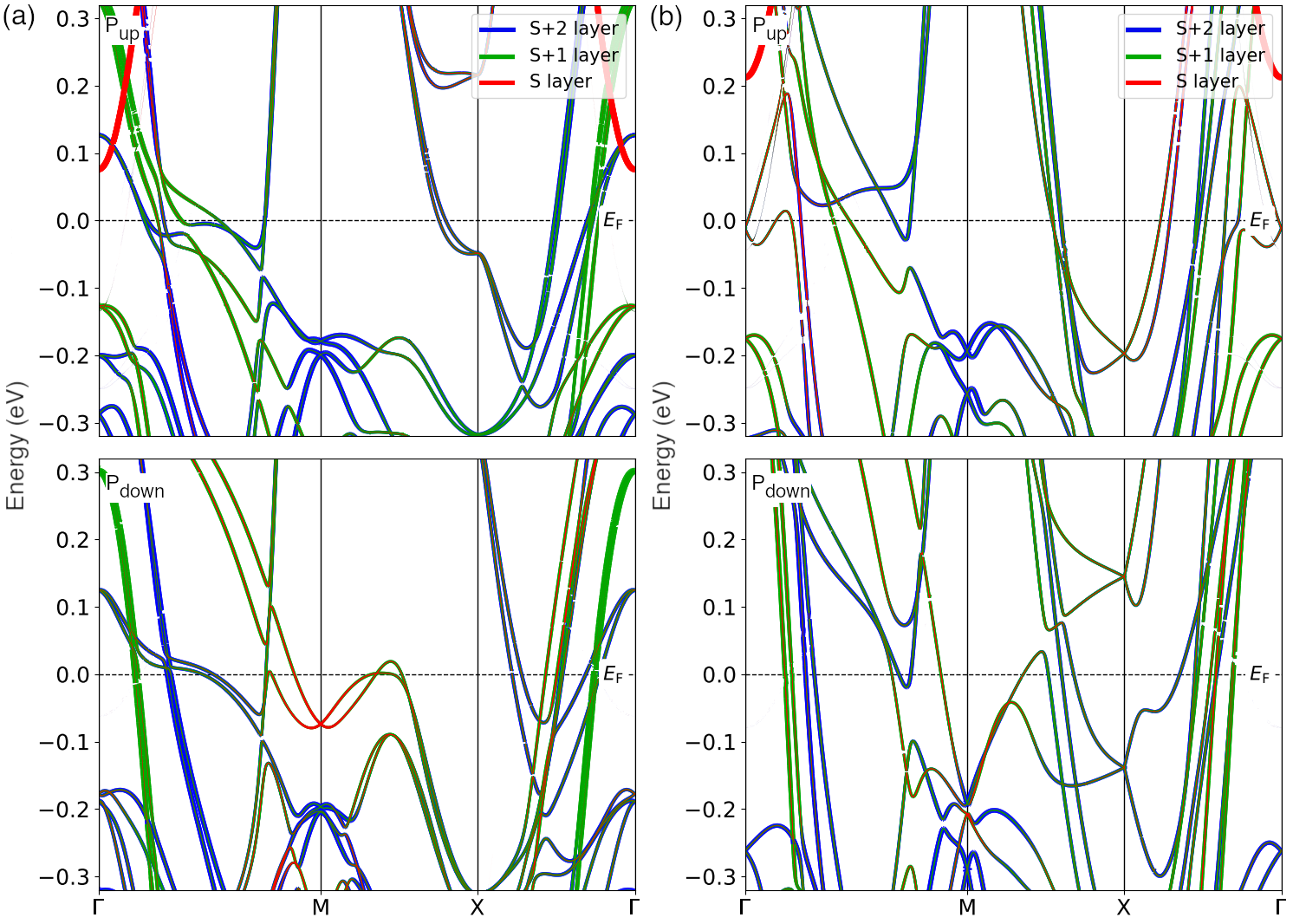}
	\caption{Band structure of (a) 2ML Pd/PTO and (b) 2ML Pt/PTO calculated for polarization P$\uparrow$ (P$\downarrow$) shown in the upper (lower) panel.
           Red, green and blue lines denote the bands emerged from the layers S, S+1  and S+2, respectively.
           The linewidth is proportional to the state density $n_i(E)$ of each branch at given energy.} 
	\label{fig9}
\end{figure*}

It is well known that the density functional theory (DFT), within both, the GGA and LDA,
underestimates the insulating band gap of ferroelectric PTO, whereas DFT+$U$~\cite{anisimov1991band,anisimov1993density,solovyev1994corrected} 
improves the gap value, when the appropriate correlation parameter $U$ is applied to the 
3$d$ orbitals of Ti. For instance, the use of enormous $U$ of 7 eV increases the gap of
PTO to 2.7 eV which is, however, notably lower than the corresponding experimental value.      
Besides, the spin-polarized simulations of the multiferroic interfaces 
and tunnel junctions containing the PTO and PZT barriers~\cite{quindeau2015origin,borisov2014magnetoelectric,borisov2015spin,borisov2017multiferroic} 
showed before that DFT+$U$ induces the artificial magnetic moments up to $\sim 0.1 \mu_B$ 
on each Ti. The slab geometry and proximity to the ferromagnetic material are responsible 
for that. Thus, we used here the DFT only. The calculated band gap of  
1.5~eV keeps the insulating state for PTO and serves as a reliable starting point to simulate 
the 2DEG in $Me$/PTO. Concerning the questionable degree of electronic correlations in 
the Pd and Pt overlayers, their electronic states can be well described by DFT.

For bulk PTO, its total density of states (DOS) together with the layer-resolved 
(Pb-O and Ti-O$_2$) contributions are shown in the upper panel of Fig.~\ref{fig5}. The lower panel of  Fig.~\ref{fig5}
shows the DOS of Pd and Pt that gives an idea on how the PTO gap may be filled by the $Me$-overlayer bands.  
To analyze in details the 2DEG  and related bands, which cross the Fermi level of $Me$/PTO(001) ($Me$ $=$ Pd, Pt), 
we plot in Fig.~\ref{fig6} the layer-resolved DOS of 1ML Pt/PTO. The DOS contributions from each atom of
the layers S$-$1, S and S+1 are plotted there. For the PTO polarization P$\uparrow$, as the upper panel of Fig.~\ref{fig6}
shows, the Pt $d$  states of the 2-Pt-atom layer S+1 dominate at $E_F$ while the PTO interfacial layer S contributes 
less significantly. The layer S$-$1 (Pb-O) contributes marginally at $E_F$ indicating, therefore, the position of the former PTO band gap
which extends above $E_F$ up to $E \sim$1 eV. The lower panel of Fig.~\ref{fig6} shows that  the polarization reversal to P$\downarrow$ 
notably decreases the DOS value $n(E_F)$. This is mostly due to reduction of the Pt $d$-states whereas the S layer 
contribution changes insignificantly. Thus, we obtain a clear evidence that the PTO polarization reversal should dramatically change
the 2DEG carrier density $n(E_F)$ in $Me$/PTO(001).

Further analysis of 2DEG can be made by plotting the layer-projected band structure near $E_F$. We start from the case of 
1ML Pd/PTO shown in Fig.~\hyperref[fig8]{8\,(a)}, in two panels of which the dispersion curves $E(k)-E_F$ are plotted for each $\bf P$ 
between $-0.3$~eV and $+0.3$~eV along the high symmetry directions of the Brillouin zone (BZ) within its $k_z =0$ plane.
Importantly, all bands which cross $E_F$ belong to the Pd layer S+1 and interfacial Ti-O$_2$ layer S. For the polarization state 
P$\uparrow$, the Fermi surface (FS) seems relatively simple and includes  a small isotropic 
hole sheet seen around the BZ center
$\Gamma$ and the two double electron lenses situated in BZ along [100] between $\Gamma$ and X and along [110] between $\Gamma$ and M. 
When electric polarization of PTO is P$\downarrow$ the Fermi surface topology changes radically. This is shown in
the lower panel of Fig.~\hyperref[fig8]{8\,(a)}. The $\Gamma$-centered Fermi sheet disappears. Instead, a complicated multiple Fermi sheet object appears
around the $k$-point M $=$ (110). The multiple electron lenses, seen between $\Gamma$ and X and also between $\Gamma$ and M, 
change seriously in size and shape. Besides, few additional Fermi sheets appear here. 

Although Pd and Pt are isoelectronic metals, the band structure of 1ML Pt/PTO plotted in Fig.~\hyperref[fig8]{8\,(b)} differs from that of 1ML Pd/PTO. 
In general, the Pt overlayer S+1 provides more bands crossing $E_F$, compared to the case of Pd. For both polarizations P$\uparrow$ and 
P$\downarrow$, there are bunches of electronic branches which appear around each high-symmetry point of the BZ that may open extra channels for
carriers. Hence, the 2DEG of Pt/PTO should be considered as more suitable for the charge transfer.  The reversal of polarization indicates
numerous qualitative changes of FS, while some of them represent the electronic topological transitions
followed by valuable changes in the effective mass. Additionally, we calculated the band structure of 1ML Pt/PTO using the semi-infinite
setup of dually polar PTO(001). After relaxation the calculated band structure changes 
marginally, as compared to the use of the 5-u.c.-thick PTO.

The second $Me$ overlayer grown on PTO(001) yields more bands which cross $E_F$. The corresponding band structures of 
2ML Pd/PTO and 2ML Pt/PTO are shown in Figs.~\hyperref[fig9]{9\,(a)} and \hyperref[fig9]{9\,(b)}, respectively. As one can assume, there are
notable differences in the FS topology and effective masses emerged due to the $\bf P$-reversal. 
Here, all band structures are plotted as non spin-polarized. This is because the magnetic moments induced on Pd and Pt 
and discussed above, as the result of interplay between the open $Me$ surface and polar interface,
may be suppressed by growing the next ferromagnetic overlayer on $Me$/PTO. 
Thus, we do not focus on spin polarization of 2DEG.

The SOC effect, however, needs to be analyzed more detailed.~\cite{varignon2019electrically} 
We picked out for that 
the case of 1ML Pt/PTO, the band structure of which is plotted in Fig.~\hyperref[fig8]{8\,(b)}.
Some selected Rashba splittings of the Pt bands seen near $E_F$ are enumerated
in Fig.~\hyperref[fig8]{8\,(b)} separately for P$\uparrow$  and P$\downarrow$. This means that
the two splitted bands, which are labeled by "1" for P$\uparrow$, differ completely from the 
P$\downarrow$ case. This is simply because the polarization reversal shifts the
whole band structure far away from $E_F$ by $\sim$0.4 eV. For a better visualization of 
the Rashba SOC of 1ML Pt/PTO, we plot in Fig.~\ref{fig7} its band structure, which was 
calculated more accurately using the semi-infinite set up. One can easily detect the 
Rashba $k$-splittings  ($\Delta k$) at $E_F$. The corresponding values are collected in
Tab.~\ref{tab2}. The typical SOC splittings selected for visualization
do not exceed the value of 0.08~(1/{\AA}).  
 For each of the two splitted branches and for each polarization, we calculated their Fermi 
velocities. In general, these electron velocities range widely between 0.3$\cdot 10^4$~m/s 
and 4.1$\cdot10^4$~m/s.

However, the velocities of splitted branches differ less significantly.
In the case P$\uparrow$  and Rashba splitting  labeled by '3' in Fig.~\hyperref[fig8]{8\,(b)},
the two splitted branches show velocities of 2.8$\cdot10^4$~m/s and 3.7$\cdot10^4$~m/s, whereas
the case '2' for P$\downarrow$ gives 1.3$\cdot10^4$~m/s and 2.0$\cdot10^4$~m/s.

 \begin{figure}[!t]
     \centering
     \includegraphics[width=0.48\textwidth]{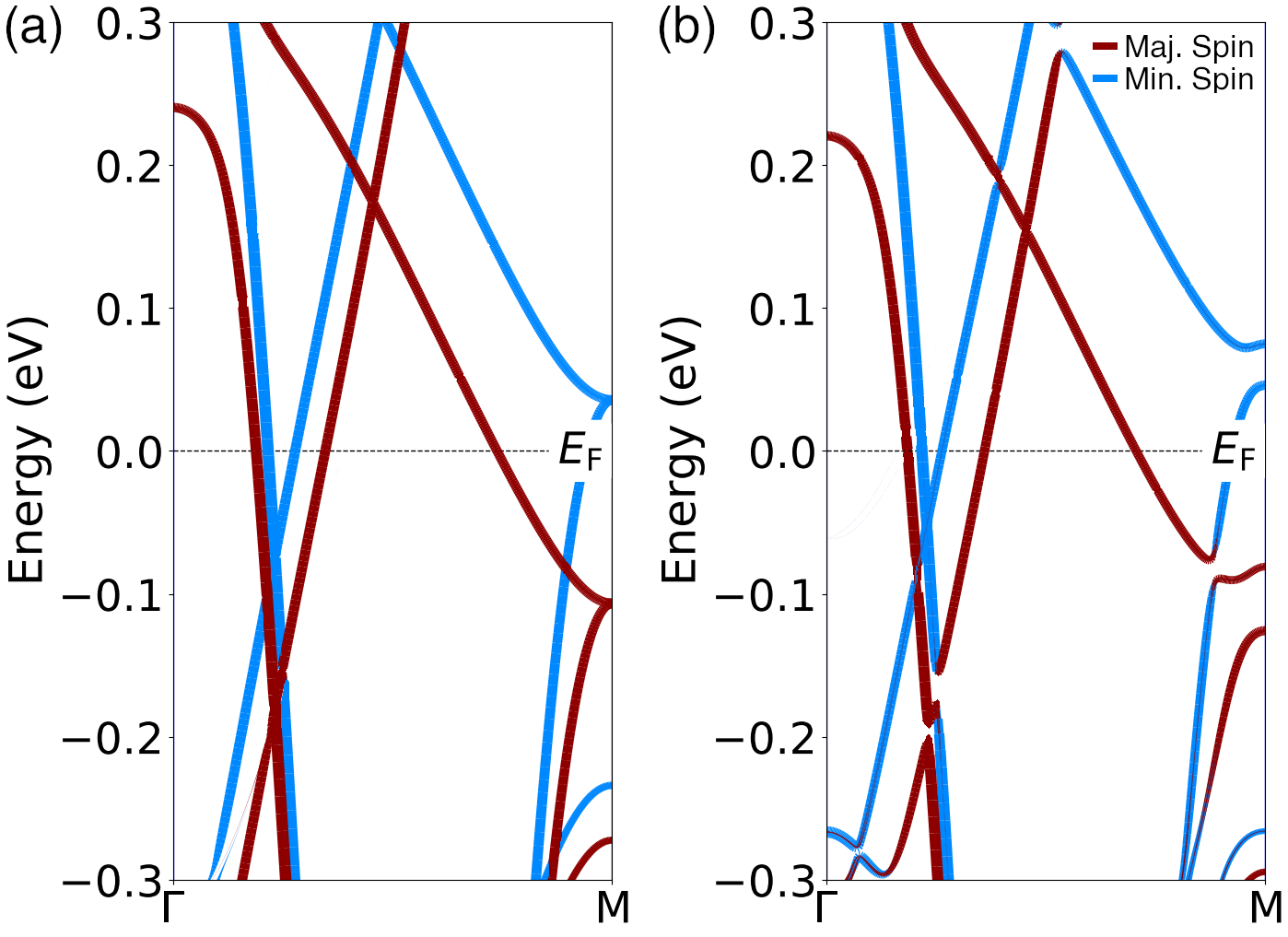}
     \caption{Spin-projected band structure of 1ML Pd/PTO (P$\downarrow$) calculated (a) without and (b) with SOC.}
     \label{fig10}
 \end{figure}

So far, we discussed the $Me$ band structure and related Rashba splittings without taking into account their spin imbalance. 
This is because robustly induced spin polarization in the $Me$ 2DEG should be considered quantitatively using ferromagnetic overlayer needed
to complete the Edelstein effect. The SOC and broken inversion symmetry
at the 2DEG interface affects the band structure and its spin polarization more seriously
than ordinary Rashba splitting. In Fig.~\ref{fig10} we plot the spin-projected band structure
of 1ML Pd/PTO calculated with and without the SOC. For a better comparison,
one BZ direction $\Gamma$--M and the case of P$\downarrow$ are shown there.
These calculations were performed allowing for a noncollinear spintexture but
keeping the spin angle $\theta =0$ that freezes the noncollinear degrees of freedom.
One can see from Fig.~\ref{fig10} that for some branches the SOC changes both topology
and spin polarization. This may affect the spin transport properties. Thus, a combination
of SOC and spin degrees of freedom within the first-principles calculations opens a way
to simulate nontrivial spin textures.

\subsection{Transport via 2DEG}

\begin{figure}[!t]
    \centering
    \includegraphics[width=0.48\textwidth]{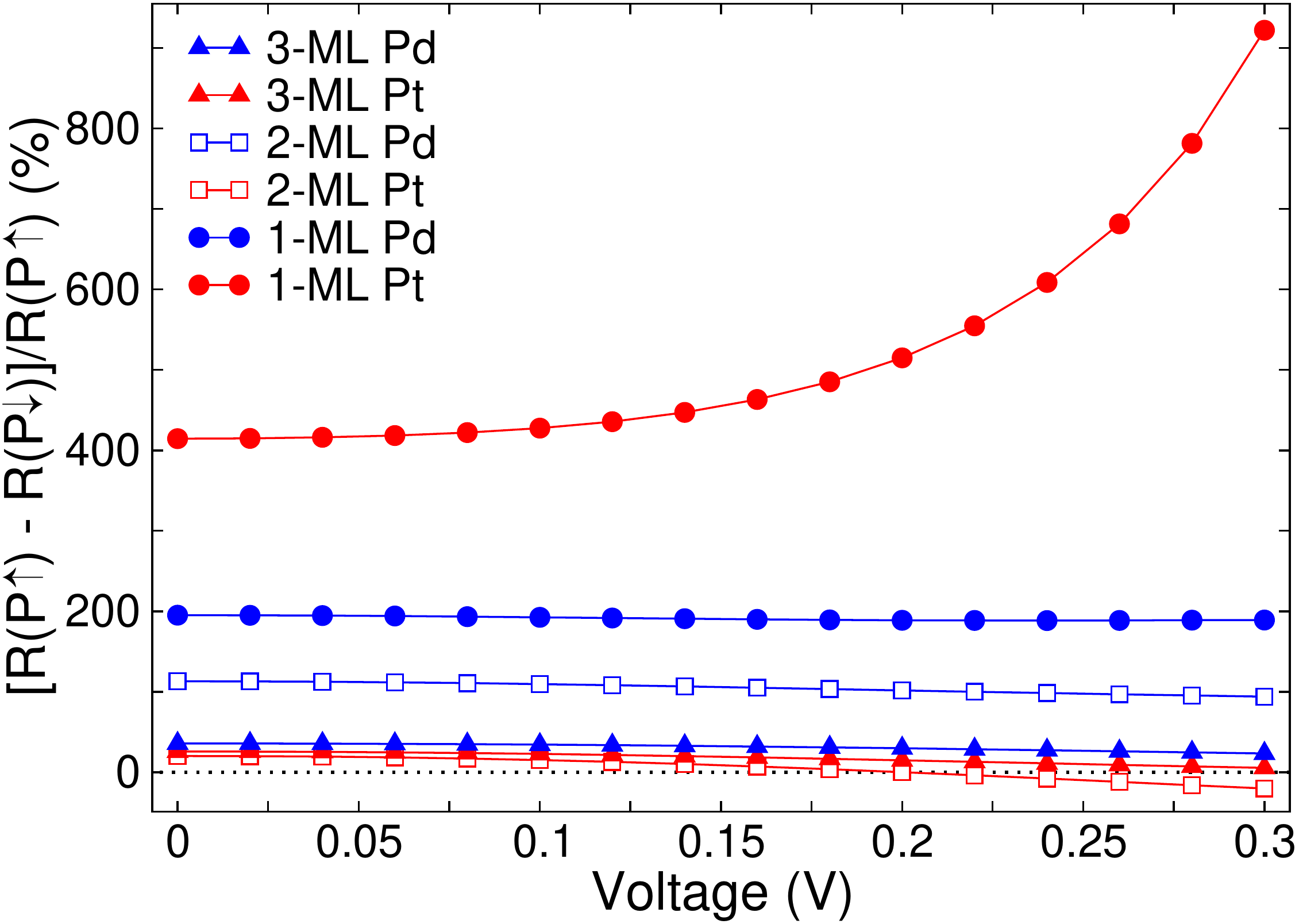}
    \caption{Difference of P$\uparrow$ and P$\downarrow$ resistance for $L$ $\cdot$ $Me$/PTO ($L$ $=$ 1, 2, 3 ML and $Me$ $=$ Pt, Pd) calculated in linear response theory.}
 \label{fig:IV}
\end{figure}

The steady-state transport calculations are performed using the \textsc{QuantumATK} package~\cite{QuantumATK}.
For this we used a ${28 \times 28 \times 3}$ $k$-point grid while we kept the density mesh cutoff of 280~Ry as well as the broadening of 25~meV and the pseudopotentials.
Afterwards we calculated the linear response current-voltage ($I-V$) characteristics by integrating the transmission spectrum for an applied bias between $0.0$~V and $+0.3$~V. 
Therefor we applied the Landauer-B{\"u}ttiker formalism~\cite{buttiker1985generalized} $ I(V) = e/h \sum_{\sigma} \int \, T^{\sigma} (E,V) \left[ f_L (E,V) - f_R (E,V) \right] \mathrm{d}E $.
The transmission coefficient $T^\sigma (E,V)$ depends on the spin $\sigma$, the energy $E$, and the applied bias voltage $V$.
For the calculation of $T^\sigma (E,V)$ we applied a {$32 \times 3$} $k$-point grid.
Details of the application of this method are provided in Refs.~\cite{brandbyge2002density,taylor2001ab,soler2002siesta}.
The calculated current-voltage ($I-V$) characteristics of $Me$/PTO were used to calculate the resistance of each system which
shows notable dependence on the PTO polarization state. Below, we discuss the case
of $L$ $\cdot$ Pd/PTO and $L$ $\cdot$ Pt/PTO ($L$ $=$ 1, 2, 3 ML), which $R$(P$\uparrow$) and $R$(P$\downarrow$) were
used to obtain [$R$(P$\uparrow$) - $R$(P$\downarrow$)] between $0.0$~V and $+0.3$~V. For each 
$Me$ $=$ Pd, Pt and each ${\bf P}$, except for 1ML Pt/PTO, the $R$ stays almost constant over the voltage range $V$
which is not surprising (see Fig.~\ref{fig:IV}). 
However, above 0.18~V, the assumption that the transmission is bias-independent seems invalid.
We anticipate that the reversal of polarization in the PTO substrate changes the
electroresistance of ultrathin metal overlayers by several percent.

\section{Summary}

We presented {\it ab initio} calculations of ferroelectrically controlled 2DEG and related 
Rashba splittings in $L$ $\cdot$ ($Me_2$)/PbTiO$_3$(001) ($Me$ $=$ Pd, Pt and $L$ $=$ 1, 2, 3 ML).
Using the reliable computational set up to mimic the semi-infinite FE 
substrate, we performed systematic calculations which show how the band structure and 
its Rashba splitting differ in the Pt and Pd overlayers and how these electronic 
features change with increasing the metal thickness and reversal of electric polarization.
From the basis of our calculations, one can conclude that platinum overlayers should be more 
suitable for the Rashba-Edelstein effect due to much stronger SOC and several electronic
topological transitions occurring under the {\bf P}-reversal.
We anticipate the 20~\% change in     electroresistance of Pt/PTO upon the {\bf P}-reversal that is a sizable value to attract extensive attention.
These findings can stimulate further experimental and theoretical studies 
of high spin-to-charge conversion efficiency.

Recently, T. Kawada {\it et al.} have shown that the lattice displacements, 
excited by the surface acoustic waves in nonmagnetic layer of  
nonmagnetic/ferromagnetic metallic bilayers grown on piezoelectric substrate LiNbO$_3$, 
including those of Pt/CoFeB and Pt/NiFe, facilitate a spin current~\cite{kawada2021acoustic}. 
The latter flows 
orthogonally to the  propagation direction of acoustic waves, while acoustic voltage 
scales with the square of the spin Hall angle of nonmagnetic layer and is proportional 
to the acoustic wave frequency.
Since PTO also possesses piezoelectricity, we suggest that Pt/PTO and Pd/PTO
covered by CoFeB or NiFe may disclose the acoustic spin Hall effect which facilitate the 
the spin current and spin–charge conversion that is the forefront of advanced spintronics.

\section{Acknowledgments}

Funding by the European Union (EFRE) via Grant No. ZS/2016/06/79307 and the Deutsche Forschungsgemeinschaft (DFG) via SFB CRC/TRR 227 is gratefully acknowledged.

\bibliography{./2DEGonPTO.bib}

\end{document}